# Ecological Notes on the Annulated Treeboa (*Corallus annulatus*) from a Costa Rican Lowland Tropical Wet Forest

Todd R. Lewis[1], Paul B.C. Grant[2], Robert W. Henderson[3], Alex Figueroa[4], and Mike D. Dunn[5]

[1]Wareham, Dorset, BH20 4PJ United Kingdom (ecolewis@gmail.com)
[2]4901 Cherry Tree Bend, Victoria, British Colombia, V8Y 1S1 Canada
[3]Milwaukee Public Museum, Milwaukee, Wisconsin 53233-1478, USA (henderson@mpm.edu)
[4]Department of Biological Sciences, University of New Orleans, New Orleans, Louisiana 70122, USA
[5]57 Ponderosa Drive, Whitehorse, Yukon Territory, Y1A 5E4 Canada

The Annulated Treeboa (*Corallus annulatus*; Figs. 1–4) is one of nine currently recognized species in the boid genus *Corallus* (Henderson et al. 2009). Its disjunct range extends from eastern Guatemala into northern Honduras, southeastern Nicaragua, northeastern Costa Rica, and southwestern Panama to northern Colombia west of the Andes (Henderson et al. 2001, McCranie 2010). It is the only species of *Corallus* found on the Caribbean versant of Costa Rica, where it occurs at elevations to at least 650 m and perhaps as high as 1,000 m (Solórzano 2004). *Corallus annulatus* occurs mostly in primary and secondary lowland tropical wet and moist rainforest (Holdridge 1967) and it appears to be genuinely rare (Henderson et al. 2001). Besides *C. cropanii* and *C. blombergi* (the latter closely related to *C. annulatus*), it is the rarest member of the genus. Aside from information on habitat and activity, little is known regarding its natural history.

In November 2001, a herpetological investigation at Caño Palma Biological Station, Tortuguero, in northeastern Costa Rica (Figs. 5 & 6) discovered the presence of *C. annulatus* from a single preserved specimen held at the biological station. Further surveys in the area detected the species in *Manicaria* swamp forest (Lewis et al. 2010) that apparently held locally common populations of the snake (Fig. 7). Further inventory and abundance surveys over the next ten years resulted in some preliminary morphometric and ecological data on *C. annulatus*.

## Study Site and Methods

Caño Palma Biological Station's climate has average daily temperatures of 26 °C (23–32 °C) and 70% RH (60–95%). The region is subject to the customary wet (September–February) and dry (March–September) seasons of the Neotropics, but is often affected by onshore weather and storms from the Caribbean Sea. The Caño Palma area has highly unpredictable annual rainfall that can exceed 6,500 mm (Lewis 2009). Over 100 cm of rain over the course of a few days, resulting in temporary flooding of the station and its grounds, is not unusual. Such abundant and serious rainfall in a known lowland catchment area creates a habitat that is very wet, botanically diverse, and unusually dominated by palms (*Manicaria saccifera*, locally known as "Palma real"), resulting in a lowland tropical wet forest on risen terrain (Myers 1990, Lewis 2009, Lewis et al. 2010).

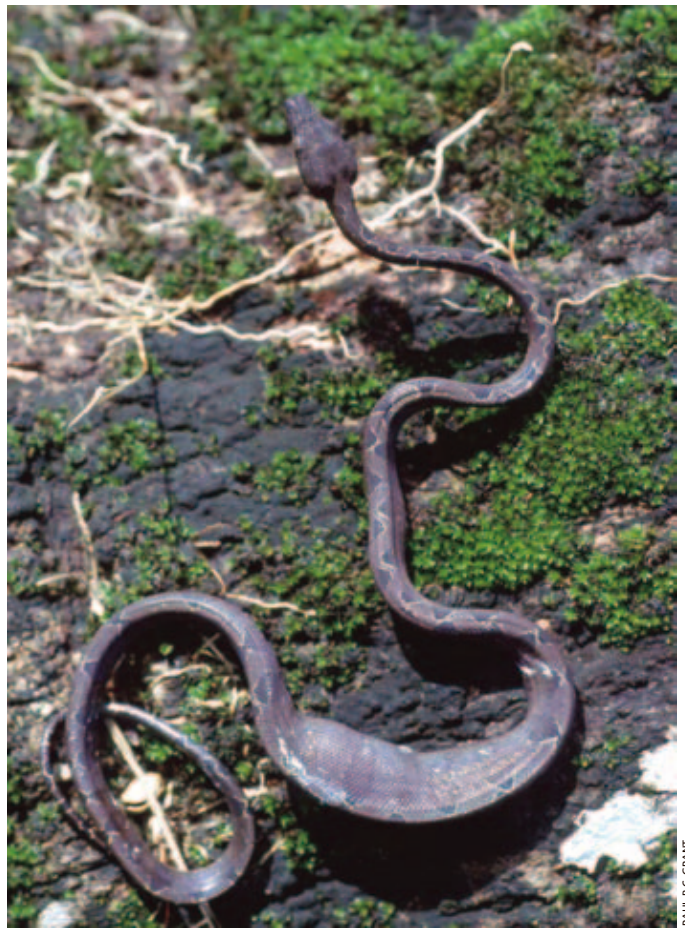

Fig. 2. *Corallus annulatus* from the Caño Palma Biological Station after consuming a bat.

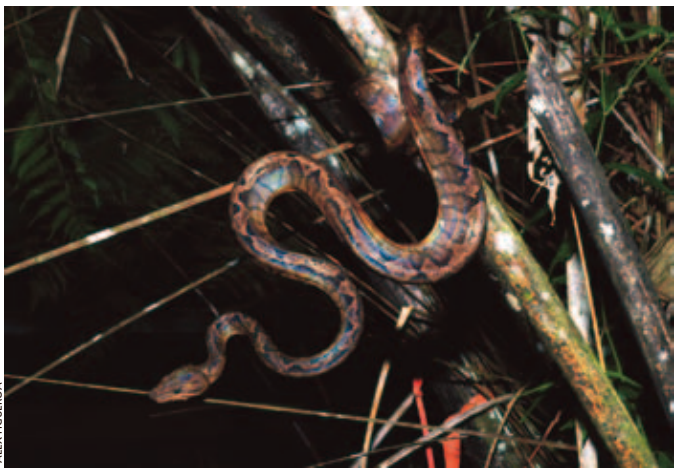

Fig. 1. *Corallus annulatus* in a palm (*Raphia taedigera*) at the Caño Palma Biological Station.

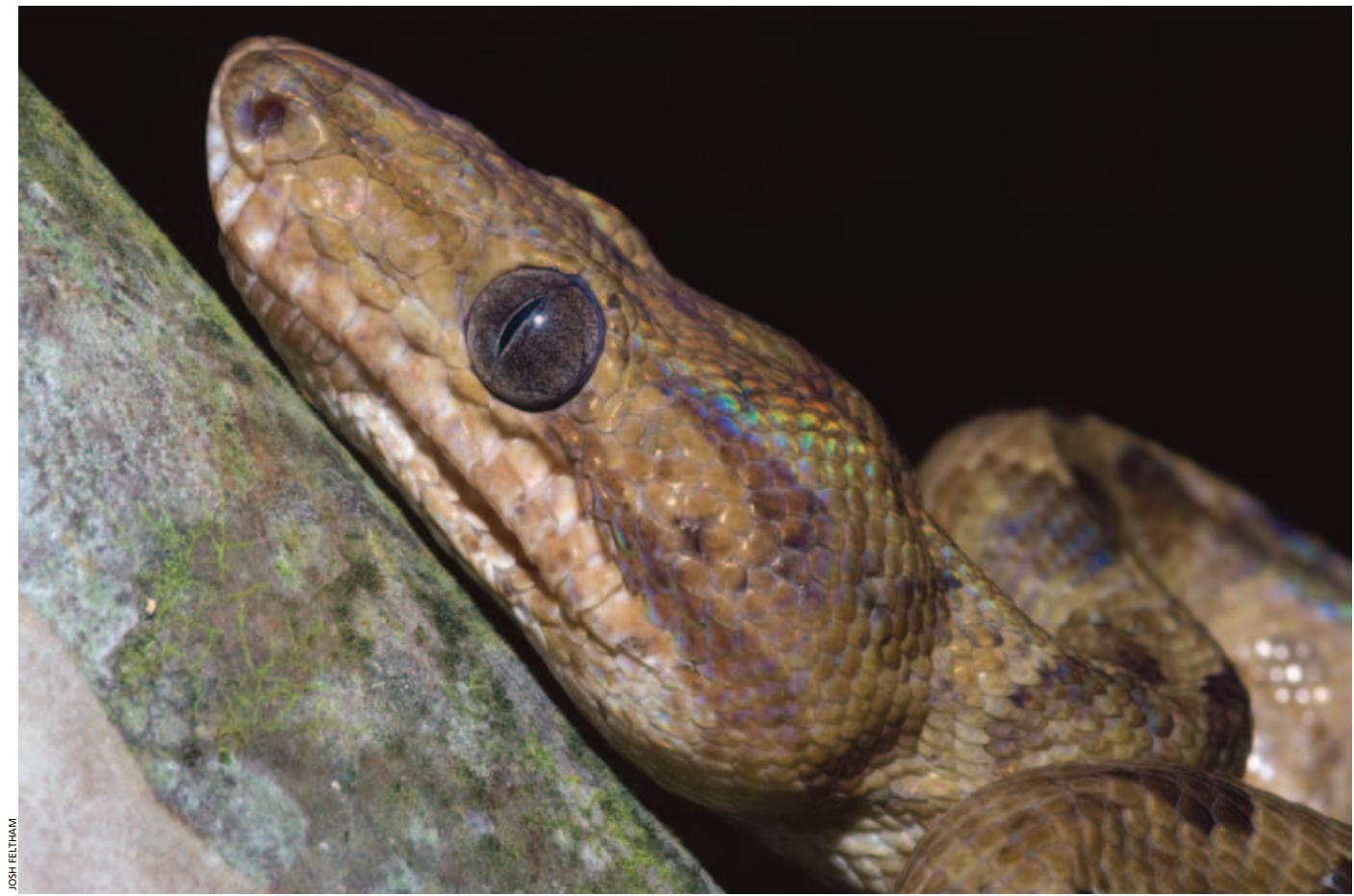

**Fig. 3.** *Corallus annulatus* from the Caño Palma Biological Station.

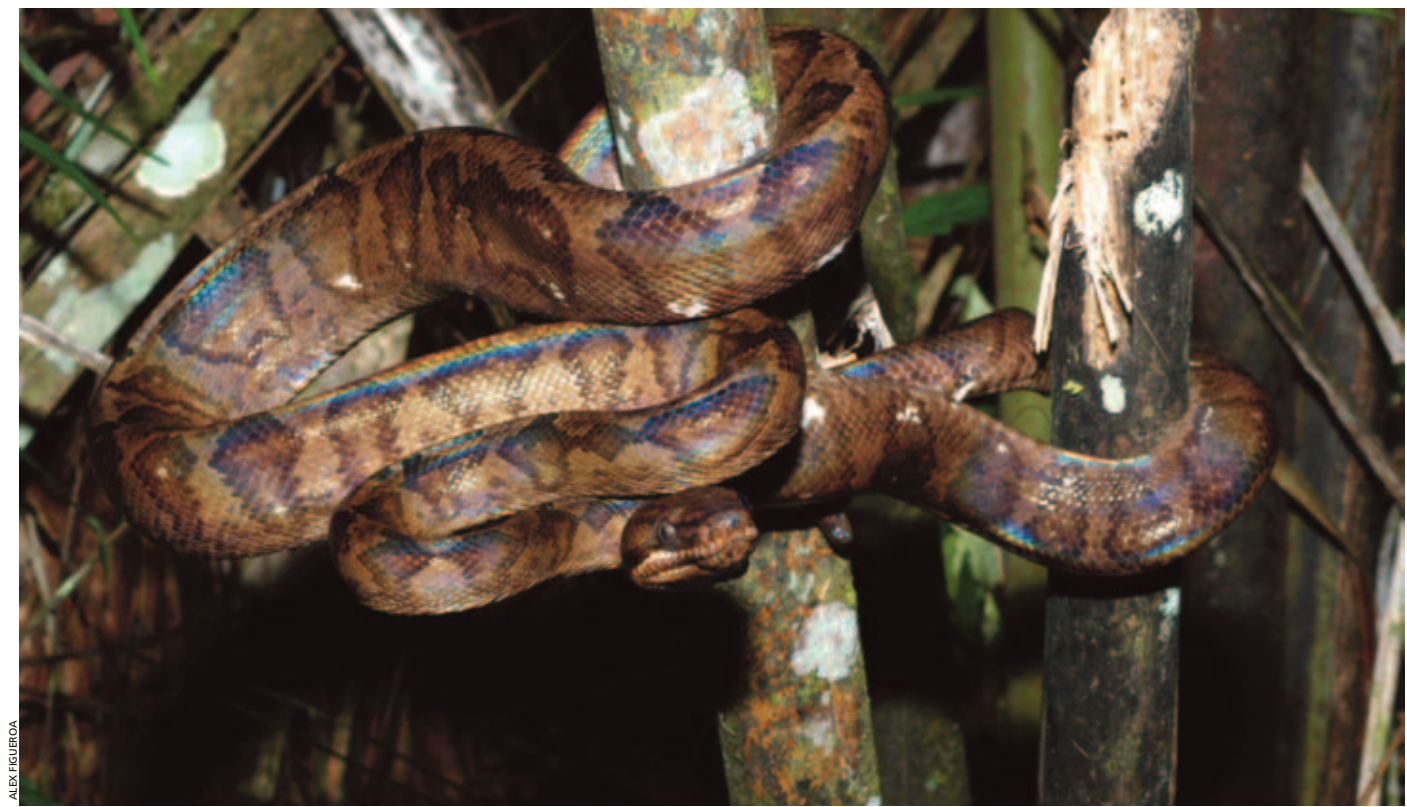

**Fig. 4.** *Corallus annulatus* in a palm (*Raphia taedigera*) at the Caño Palma Biological Station.

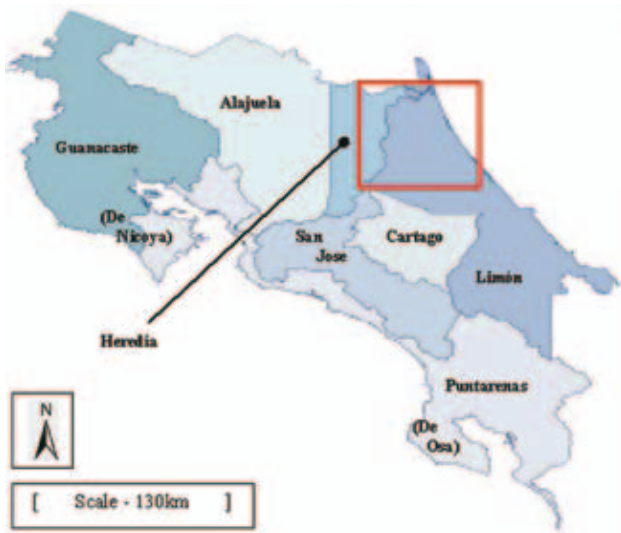

**Fig. 5.** Costa Rica's provincial and northeastern zone.

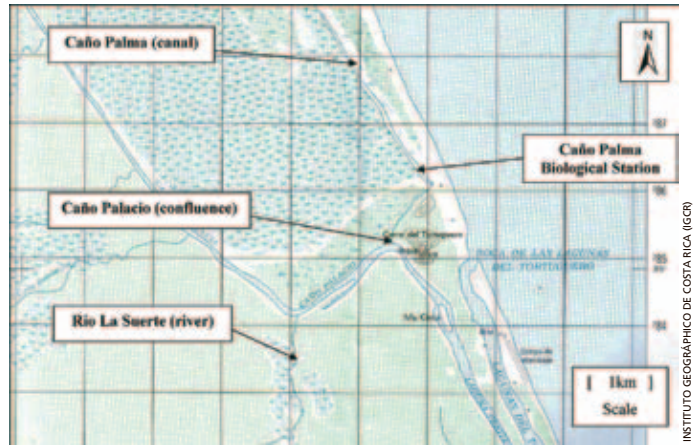

**Fig. 6.** Lithograph showing the approximate location of Caño Palma; the location of the Río La Suerte and the Caño Palacio/Caño Palma confluence.

Nocturnal visual encounter surveys (VES) (Heyer et al. 1994) were conducted on known trails and transects at Caño Palma Biological Station to locate *Corallus annulatus*. We were alerted to some additional individuals by local residents. Data recorded included morphometrics: Weight (g), snout-vent length (SVL), total length (TL); as well as sex, habitat, diet, and perch height (PH), where possible. Survey and morphometric measurements were carried out under research permits granted by the Ministerio del Ambiente Energía Telecomunicaciones (MINAET) under the Sistema Nacional de Áreas de Conservación (SINAC) investigation program. In the interest of conserving the species, and despite a lack of preserved specimens in known collections globally, no specimens were collected. Spearman Rank correlation coefficients ($r_s$) conducted in Statistica™ Ver. 7 (StatSoft, Inc., Tulsa Oklahoma) were used to investigate morphological characteristics.

### Results and Discussion

Over a span of ten years, we captured ten female and ten male *C. annulatus* (Table 1), but observed 26 in total. Based on size data, 17 of the snakes were adults and three were juveniles. Mean SVL of males was 749.6 ± 81.1 mm (198–1,020 mm) and for females 700.3 ± 107.8 (180–1,434 mm) (Fig. 8). We found a female of 1,434 mm SVL and 1,544 mm total length, very close in size to the largest known individual, also a female, of 1,447 mm SVL (1,725 mm total length) from Guatemala (Smith and Acevedo 1997). Mass varied between 27 and 448 g. Total length, SVL, and mass were all positively correlated ($r_s$ = 0.789, 0.955, 0.913; $P < 0.05$).

**Table 1.** Morphometric and habitat details of *Corallus annulatus* at Caño Palma Biological Station, 2001–2011 (SVL and total length in mm, mass in g, perch height in cm; NR = not recorded).

| Date | Sex | Age | Color | SVL (TL)/Mass | Perch Height | Habitat | Prey |
|---|---|---|---|---|---|---|---|
| 01 May 2001 | F | Adult | orange | 580 (654)/57 | 250 | *Pentaclethra macroloba* | — |
| 16 Aug 2001 | M | Adult | gray | 593 (800)/65 | 850 | *Bambusa vulgaris* | — |
| 21 Aug 2001 | M | Adult | orange | 992 (1,306)/172 | 300 | *Manicaria saccifera* | — |
| 12 Oct 2001 | M | Adult | light gray | 825 (1,210)/168 | 900 | *Bambusa vulgaris* | — |
| 28 Oct 2001 | F | Adult | gray | 600 (720)/56 | 380 | *Citrus aurantiun* | — |
| 12 Jan 2002 | F | Juvenile | black | 180 (270)/27 | 240 | Boat dock | *Rhynchonycteris naso* |
| 16 Nov 2002 | M | Adult | gray/orange | 672 (990)/96 | 850 | *Bambusa vulgaris* | *Wilsonia canadensis* |
| 29 Nov 2002 | F | Adult | orange | 590 (660)/44.5 | 550 | *Bambusa vulgaris* | — |
| 07 Jan 2003 | M | Adult | orange | 900 (1,100)/157 | 180 | *Bambusa vulgaris* | — |
| 15 Jul 2003 | M | Adult | orange | 512 (584)/36 | 200 | *Manicaria saccifera* | *Rhynchonycteris naso* |
| 28 Nov 2003 | M | Juvenile | gray/orange | 198 (287)/38 | 280 | *Manicaria saccifera* | — |
| 06 Dec 2004 | M | Adult | orange | 1,020 (1,200)/165 | 30 | *Manicaria saccifera* | — |
| 27 Jan 2005 | M | Adult | orange | 910 (1,250)/170 | 500 | *Manicaria saccifera* | — |
| 03 Feb 2005 | F | Adult | orange | 780 (970)/237 | 250 | *Manicaria saccifera* | — |
| 15 Feb 2010 | F | Juvenile | gray | 392 (472)/NR | NR | Pond shrub | — |
| 17 Jul 2010 | M | Adult | orange | 874 (1,008)/NR | NR | Rancho building | Unidentified bat |
| 15 Sep 2010 | F | Adult | orange | 1,434 (1,544)/448 | 120 | *Zygia latifolia* | — |
| 28 Dec 2010 | F | Adult | gray/orange | 968 (1,077)/282 | 120 | *Vochysia ferrugenia* | — |
| 10 Jun 2011 | F | Adult | gray/orange | 850 (1,013)/141 | 150 | *Mangifera indica* | — |
| 15 Jun 2011 | F | Adult | orange | 629 (738)/90 | 100 | *Heliconia* sp. | — |

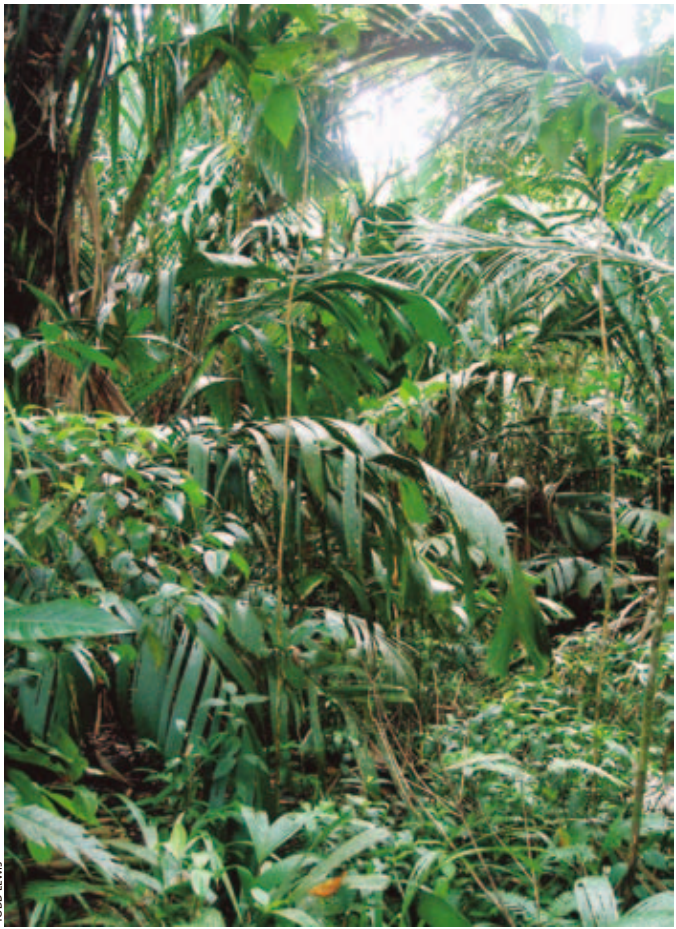

**Fig. 7.** Typical *Manicaria* palm swamp forest, dominated by Palma Real (*Manicaria saccifera*), at the northern section of the Caño Palma Biological Station's forest.

Fifteen of the boas were orange to cinnamon in dorsal coloration, with brown/orange saddle markings. Four were a slate-gray to brown with dark crossbands and saddle markings. Only one individual was black, but we do not think it was melanistic or anerythristic due to the fact that it showed partial saddle markings and lighter markings on the underside. Color did not correlate significantly with morphology ($r_s$ = -0.464; $P > 0.05$) and did not relate to gender.

Habitats varied considerably, but we found no significant correlation between habitat type, morphology, or color ($r_s$ = 0.078–0.248; $P > 0.05$). Six individuals were found in primary *Manicaria* swamp forest and two were found in human-made structures within the biological station. A favored area (five individuals) was a riparian stand of mature, introduced Bamboo (*Bambusa* sp.) located in the garden area of the biological station. Here we found a boa that had swallowed a Canadian Wilson's Warbler (*Wilsonia canadensis*), a known winter migrant to the region. Clearly, *C. annulatus* is able to utilize habitats in natural and human-altered situations and, like several other species of *Corallus* (*C. cookii*, *C. grenadensis*, *C. hortulanus*), does not hesitate to exploit man-made structures (Henderson, 2002, Powell et al. 2007).

In addition to the warbler, we observed a small (180 mm SVL) *C. annulatus* feeding on a Brazilian Long-nosed Bat (*Rhynchonycteris naso*) (Lewis et al. 2009). In 2010, one of us (MD) observed a boa consuming a bat in a Rancho-style building at the biological station. This bat was probably also a *R. naso*, as this species was often found in the Rancho, but its identity was not confirmed. To the best of our knowledge, small rodents were previously the only documented prey for wild *C. annulatus* (Henderson et al. 1995). Based on data for other species of *Corallus* (i.e., *C. grenadensis*, *C. hortulanus*, and *C. ruschenbergerii*; Henderson 2002 and papers cited therein), treeboas with SVLs of ~900 mm and longer are largely mammal predators and characteristically exhibit an ambush-foraging strategy, wherein they position themselves low in vegetation above a trail perceived to be used by small mammals. These ambush sites are usually lower in the vegetation when compared to the heights at which smaller boas actively forage for nocturnally quiescent prey (i.e., sleeping lizards and birds). The six *C. annulatus* in our sample with SVLs ≥900 mm (mean = 1037.3 ± 81.6 mm; range = 900–1,434 mm) were encountered at a mean perch height of 208.3 ± 68.7 cm (30–500 cm). In contrast, boas with SVLs <900 mm (mean = 584.1 ± 61.2 mm; 180–874 mm) had a mean perch height of 416.7 ± 84.9 cm (100–900 cm), or twice as high as larger individuals, although perch heights were not significantly correlated with SVL ($r_s$ = -0.198; $P > 0.05$), probably attributable to our small sample sizes. Based on these data, we suspect that *C. annulatus*, like other species in the genus, is largely a small mammal predator.

Activity in *C. annulatus* was usually minimal upon encounter. Nineteen of the individuals captured, and 23 of the total seen, were found at night, coiled on branches before moving slowly away in an upward direction when approached. Most were detected by their lighter-colored underside and telltale reflective orange eye-shine. Upon capture, many individuals attempted a swift and occasionally successful bite; other defensive behaviors included assuming a balled-up posture, voiding the cloaca, and tail vibration. Nearly all boas appeared to be healthy, with only two possessing small skin lesions or harboring external parasites such as ticks (Acari).

Despite the size of *C. annulatus* and the conspicuous red-orange eye-shine in the beam of a headlamp, only 26 individuals were encountered over a 10-year period, further attesting to the rarity of the species. In contrast, one can observe several other snake species more frequently at Caño Palma Biological Station. For example, 29 Eyelash Vipers (*Bothriechis schlegelii*), 22 Allen's Coral Snakes (*Micrurus alleni*), 13 Blunt-headed Treesnakes (*Imantodes cenchoa*), 13 Clouded Snail-suckers (*Sibon nebulatus*), 4 Ringed Snail-eaters (*Sibon annulatus*), and 4 Red Coffee Snakes (*Ninia sebae*) were observed over a four-month span (AF, unpublished data). We hope that further data collection, only possible over many years of fieldwork, will reveal more of the elusive ecology and natural history of *C. annulatus*. That the species warrants special attention by conservation biologists is supplemental to the more urgent requirement to ensure that both the primary and secondary forests where the species occurs are, or remain, protected for future generations to gather data.

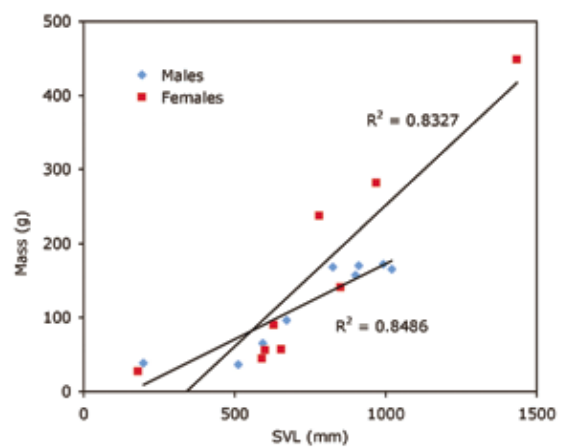

**Fig. 8.** Plot of snout-vent length (SVL) and mass for male and female *Corallus annulatus* from the Caño Palma Biological Station, 2001–2011.

### Acknowledgements

We thank Ana Maria Monge, Javier Guevara, and Carlos Calvo (MINAET) for assistance with licensing and support for the work conducted in the Barra del Colorado Wildlife Refuge and Tortuguero National Park, Costa Rica. We thank Tom Mason, Gabriel David, Daryl Loth, Ross Ballard, and Josh Feltham for informative discussions pertaining to the species. The Canadian Organization for Tropical Education and Rainforest Conservation kindly permitted the long-term investigation on its property. We also thank the late Peter Stafford, who, despite never seeing this study come to fruition, eagerly supported it from conception.

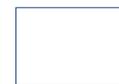